\documentclass[twocolumn,prl,showpacs]{revtex4}

\usepackage{amsmath,amssymb}
\usepackage{graphics}
\usepackage{epsfig}
\usepackage{color}

\begin{document}


\title{Collisional Aspects of 
Bosonic and Fermionic 
Dipoles in Quasi-Two-Dimensional Confining Geometries}

\author{Jos\'e P. D'Incao}
\affiliation{Department of Physics and JILA, University of Colorado,
Boulder, CO 80309-0440, USA}
\author{Chris H. Greene}
\affiliation{Department of Physics and JILA, University of Colorado,
Boulder, CO 80309-0440, USA}

\begin{abstract}
Fundamental aspects of ultracold collisions between identical bosonic or fermionic dipoles are studied under 
quasi-two-dimensional (Q2D) confinement. 
In the strongly dipolar regime, bosonic and fermion species are found to share important collisional 
properties as a result of the confining geometry, which suppresses the inelastic rates irrespective of the quantum statistics obeyed.
A potential negative is that the confinement causes dipole-dipole resonances to be extremely narrow, which could make
it difficult to explore Q2D dipolar gases with tunable interactions. Such properties are shown to be universal, and a simple 
WKB model reproduces most of our numerical results.
In order to shed light on the many-body behavior of dipolar gases in Q2D we have analyzed the scattering amplitude
and developed an energy-analytic form of the pseudopotentials for 
dipoles. For specific values of the dipolar interaction, the pseudopotential coefficient can be tuned
to arbitrarily large values, indicating the possibility of realizing Q2D dipolar gases with tunable interactions.
\end{abstract}
\maketitle 

Extensive experimental and theoretical efforts have recently been devoted to explore
the production of dipolar gases and to uncover novel quantum phases resulting from the anisotropic nature and 
long-range character of the dipolar interaction \cite{Baranov}. The recent experimental realization of a dense sample of 
ultracold ground-state KRb molecules \cite{RbK} has opened up ways to realize several new phenomena, 
ranging from ultracold chemistry, condensed matter physics and quantum information 
\cite{RbK,Krems,ManyBodyQInfo}. A crucial step towards realizing such ideas is to understand the collisional properties
of dipoles at ultralow temperatures. In fact, recent theoretical work has proposed a classification of ground-state 
molecules in terms of their chemical reactivity \cite{PolarMolClass}. 
A number of such molecules are expected to be extremely reactive
while others have  energetically forbidden decay channels. Although reactive molecules are 
great candidates for studying chemical dynamics \cite{RbK,Krems}, such processes 
limit the lifetime of the gas. The introduction of a quasi-two-dimensional confinement (Q2D) 
adds a new handle for controlling the physics of the problem \cite{Q2DPolar}, which drastically changes this scenario \cite{Ticknor,Polar2D} and 
ultimately leads to improved stability. This should in turn enable an exploration of various many-body phenomena. 

In this paper, we explore the effect of Q2D confinement on collisions of bosonic and fermionic dipolar species.
We have found that in the strong dipolar regime the ultracold scattering properties
of Q2D dipoles do not depend on the details of the short-range interactions. 
The effective interactions are also characterized by calculating the Q2D pseudopotential coupling constants.  
Our analysis shows that at {\em any} finite collision energy there exists a value of the dipolar interaction at which 
the coupling constant diverges, a phenomenon that can be traced back to the Ramsauer-Townsend effect. 
Although our present analysis does not address the issues of implementing our pseudopodential formalism to the 
many-body problem, we believe that our findings can shed light on the possible ways of controlling the interactions  
in a dipolar gas.

In order to study dipolar collisions in Q2D, an adiabatic separation of the radial and angular motion is introduced.
This expresses the wavefunction as 
$\Psi(\vec{r})= e^{i m\phi}\sum_{\nu}r^{-1}{F_{\nu}(r)}\Phi_{\nu}(r;\theta)$, 
where $r$, $\theta$, and $\phi$, are the spherical radius and angles, respectively, and $m$ is 
the angular momentum projection. Here, $\nu$ is the channel index,
$F_{\nu}$ is the $\nu$-th radial wave function and $\Phi$ is the channel function.
In the adiabatic representation, the Schr{\"o}dinger equation reduces to a simple system 
of ordinary differential equations given (in atomic units) by, 
\begin{eqnarray}
\left[-\frac{1}{2\mu}\frac{d^2}{dr^2}+W_{\nu}(r)-E\right]F_{\nu}
+\sum_{\nu'\ne\nu} W_{\nu\nu'}(r) F_{\nu'}=0,
\label{radeq}
\end{eqnarray}
where $\mu$ is the two-body reduced mass, $E$ is the total energy, 
$W_{\nu\nu'}=-\left(P_{\nu\nu'}{\partial}/{\partial r}+Q_{\nu\nu'}\right)/{2\mu}$,
with $P_{\nu\nu'}=\langle{\Phi_{\nu}}|{\partial}/{\partial r}|{\Phi_{\nu'}}\rangle$ and 
$Q_{\nu\nu'}=\langle{\Phi_{\nu}}|{\partial^2}/{\partial r^2}|{\Phi_{\nu'}}\rangle$, are
the nonadiabatic couplings which drive inelastic transitions,
and $W_{\nu}=U_{\nu}-Q_{\nu\nu}/{2\mu}$ are the adiabatic potentials supporting bound and quasi-bound
states. The adiabatic potentials and channel functions are obtained by solving 
eigenvalue equation for the angular motion: 
\begin{eqnarray}
\left[
-\frac{1}{2\mu r^2}\left(\frac{1}{\sin\theta}\frac{\partial}{\partial\theta}\sin\theta\frac{\partial}{\partial\theta}
-\frac{m^2}{\sin^2\theta}\right)+v_{sr}(r)~~~~~~~~\right.\nonumber\\
\left. +\frac{d_{\ell}}{\mu}\frac{1-3\cos^2\theta}{r^3}
                + \frac{r^2\cos^2\theta}{2\mu a_{ho}^4}-U_{\nu}(r)\right]
\Phi_{\nu}(r;\theta)
=0,
\label{adeq}
\end{eqnarray}
\noindent
solved for fixed values of $r$ and in concert with the proper bosonic and fermionic boundary conditions at $\theta=\pi/2$. 
In the equation above, $v_{sr}(r)=D{\rm sech}^2(r/r_{0})$ is the short-range isotropic potential, where $r_{0}$
is the characteristic range of the interactions (similar to the van der Waals length), and the next two terms 
are the anisotropic dipole-dipole and Q2D confinement potentials, respectively. The dipolar interaction and 
confinement introduce two new important length scales into the system, namely,
the ``dipole length'' $d_{\ell}=\mu d_{m}^2$, where $d_{m}$ is the dipole moment in a.u., and the reduced-mass
oscillator length $a_{ho}^2=\mu w_{ho}$, where $w_{ho}$ is the harmonic trap frequency experienced by each molecule in the $\hat{z}$ direction.

\begin{figure}[htbp]
\includegraphics[width=3.2in,angle=0,clip=true]{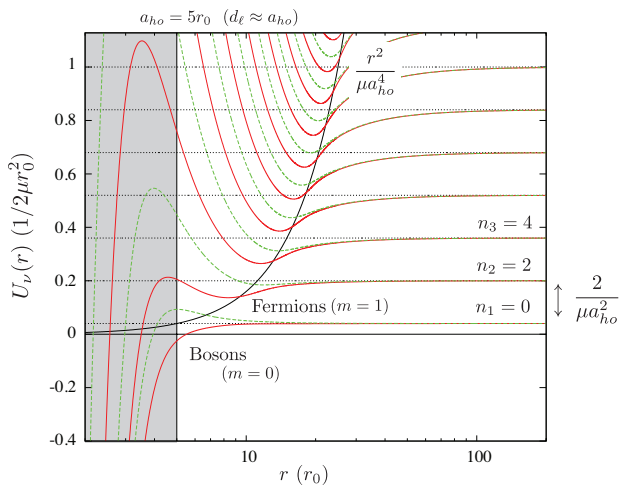}
\caption{(color online). Adiabatic potentials for Q2D bosonic (solid) and fermionic (dashed)  dipoles 
 (see text). 
\label{Fig1}} 
\end{figure}

In Figure~\ref{Fig1} we show a set of adiabatic potentials obtained by solving Eq.~(\ref{adeq})
for the lowest bosonic ($m=0$) and fermionic ($m=1$) symmetries with even $z$-reflection parity. For distances $r\lesssim a_{ho}$, the adiabatic potentials are 
mainly controlled by the short-range interactions ($v_{sr}$ and dipolar).
As $r$ increases the confinement becomes important and the transition between 3D and 2D physics is
characterized by the series of avoided crossings along the values of the averaged confinement potential 
$r^2/\mu a_{ho}^4$ (see the solid curve in Fig.~\ref{Fig1}). 
For ultracold collisions, however, the scattering properties are mainly determined by the asymptotic forms of the 
adiabatic potentials which we found, for $r\gg r_{ho}=(4a_{ho}^4d_{\ell})^{1/5}$, to be given by
\begin{eqnarray} 
&&{W_{\nu}(r)}\approx E_{ho}^{(\nu)}+\frac{m^2-1/4}{2\mu r^2}+\frac{d_{\ell}}{\mu r^3}.
\label{asypot}
\end{eqnarray} 
Here $E^{(\nu)}_{ho}=(n_{\nu}+1/2)/\mu a_{ho}^2$ is the oscillator energy level and $n_{\nu}$ is the principal quantum number. 
Notice that in contrast to the 3D case \cite{Polar3D} the dipolar interaction at large 
$r$ is repulsive $1/r^3$ for {\em both} bosonic and fermionic species. 
This fact has important consequences for the collisional properties of dipoles in Q2D, as 
we discuss below.

\begin{figure*}[htbp]
\includegraphics[width=6.9in,angle=0,clip=true]{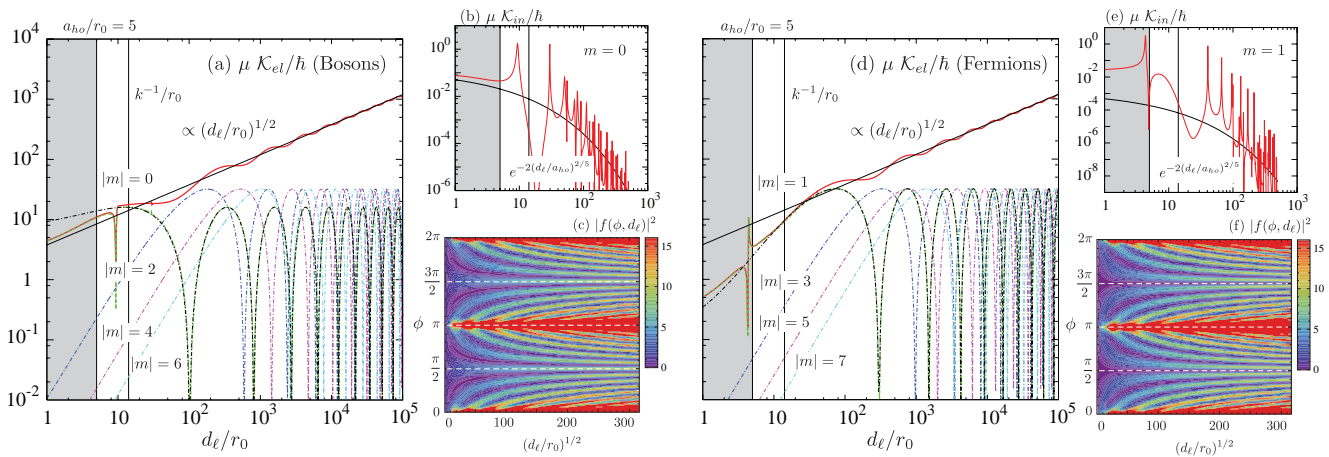}
\caption{(color online). Elastic and inelastic rates for bosonic, (a) and (b), and fermionic, (d) and (e), dipoles as a function
of $d_{\ell}$ for $a_{ho}=5r_{0}$ and $k=0.071r_{0}^{-1}$. 
In the strongly dipolar regime, $k d_{\ell}\gg1$, the elastic rates approaches to their semi-classical expectation \cite{Ticknor} [solid straight line
in (a) and (d)] and the inelastic rates are strongly suppressed [(b) and (e)]. (c) and (f) shows the scattering amplitude displaying anisotropy 
effects (see text).}
\label{Fig2}
\end{figure*}

Having determined the adiabatic potentials and couplings, we calculate elastic and inelastic rates by solving 
Eq.~(\ref{radeq}) and matching our numerical solutions to the proper cylindrical asymptotic solutions to obtain the 
corresponding $T$-matrix and the collision rate coefficients \cite{Polar2D},
\begin{eqnarray}
{\cal K}_{el}=\frac{2\hbar}{\mu}\sum_{m, f}\Delta_{m}|T^{(m)}_{i\leftarrow i}|^2,~~~
{\cal K}_{in}=\frac{2\hbar}{\mu}\sum_{m, f}\Delta_{m}|T^{(m)}_{f\leftarrow i}|^2,
\label{rates}
\end{eqnarray}
where $\Delta_{m}=2-\delta_{m,0}$ and $i$ and $f$ label the initial and final collision channels. Here, we study collisions from
the lowest transverse mode ($n_{1}=0$ in Fig.~\ref{Fig1}) and we have added an artificial deeply bound channel (with adjustable
coupling) to incorporate the effects of inelastic collisions relevant for reactive ground-state polar molecules \cite{PolarMolClass}. 

In general, a calculation of elastic and inelastic rates in the strong dipolar regime, $k d_{\ell}\gg1$, where
$k^2=2\mu (E-E_{ho}^{(0)})$, requires the inclusion of a large number of partial waves in Eq.~(\ref{rates}). Here, however, 
we have found that the rates in this regime assume a universal form, allowing for a simple WKB model for the phase-shifts
that tremendously economizes the calculations. We compare our numerical results for the elastic rates for $m=0$ and $1$ 
with the WKB model in order to demonstrate this universality and present the calculation for the total rates [Eq.~(\ref{rates})] 
for values of $m$ up to 200 using our WKB result. The topology
of the adiabatic potentials (Fig.~\ref{Fig1}) allows us to directly apply the WKB phase-shift formula derived
in Ref. \cite{Berry}. Here, however, we write the phase-shift as a sum of two terms,
\begin{eqnarray}
\delta_{m}(k)=\delta^{lr}_{m}(k)+\delta^{sr}_{m}(k),\label{WKB}
\end{eqnarray}
where $\delta^{lr}_{m}=\int_{r^{c}_{m}}^{\infty}\left[K_{m}(r)-k\right]dr+\frac{\pi}{2}m-kr^{c}_{m}$ is the long-range phase-shift, 
$r^c_{m}$ being the classical turning point \cite{rc}, and 
$\delta^{sr}_{m}=\tan^{-1}[\frac{1}{4}e^{-2\gamma_{m}}\tan\Phi^{sr}_{m}]$ is the short-range phase-shift. 
Here $K_{m}(r)=[k^2-(2d_{\ell}+m^2 r)/r^3]^{1/2}$, $e^{-2\gamma_{m}}=\exp[-2\int_{r_{ho}}^{r^{c}_{m}}|K_{m}(r)|dr$].
In our formulation, the only non-universal part of the phase-shift comes from $\delta_{m}^{sr}$ through the phase 
$\Phi_{m}^{sr}$ accumulated for $r<r_{ho}$. We have determined, however, that for $kd_{\ell}\gg1$, the phase $\delta_{m}^{sr}$ is
exponentially small [$\gamma_{m}\propto (d_{\ell}/a_{ho})^{2/5}$], due to the 
$1/r^3$ barrier in the entrance channel [Eq.~(\ref{asypot})], except in an extremely narrow region near
a dipole-dipole resonance \cite{DipolarPhase}. As a result the scattering problem becomes 
universal, depending {\em only} on the long-range physics encapsulated in $\delta_{m}^{lr}\equiv\delta_{m}^{lr}(k,d_{\ell})$.
The fact that, in Q2D, dipole-dipole resonances are extremely narrow, seems likely to make the exploration of Q2D dipolar gases with tunable interactions prohibitively difficult. However, as we will show later, tunability is still possible, although instead due to the physics of the Ramsauer-Townsend effect.

Figure~\ref{Fig2} demonstrates some of these points through our numerical calculations (for $m=0$ and $1$) and WKB results for the partial rates
 ${\cal K}^{m}_{el}=(8\hbar/\mu)\Delta_{m}\sin^2\delta_{m}(k)$ and ${\cal K}^{m}_{in}$ for
$a_{ho}/r_{0}=5$ and $k=0.071/r_{0}$ (vertical lines in Fig.~\ref{Fig2}).
Figs.~\ref{Fig2}~(a) and (d) compare our numerical results for $m=0$ and $m=1$ (thick dashed lines)
with the WKB results from Eq.~(\ref{WKB}) (thin dot-dashed lines). Extremely good agreement is found for 
$k d_{\ell}>1$, but for $k d_{\ell}<1$, this agreement deteriorates and dipole-dipole resonances \cite{DipolarPhase} becomes visible.
In fact, throughout the entire range of values of $d_{\ell}$ there exist dipole-dipole resonances, 
but as we mentioned above, they are extremely narrow and only become visible in the inelastic 
rates [Figs.~\ref{Fig2}~(b) and (e)].
Interestingly, all partial rates oscillate as $d_{\ell}$ increases, as a result of the Ramsauer-Townsend effect.
Evidently, the total rates will not display actual zeros since the partial rates are out of phase.
Nevertheless, the total rates, shown in Figs.~\ref{Fig2}~(a) and (d) as solid lines, oscillate around the
semi-classical result of Ref.~\cite{Ticknor} (solid straight line). Still, such oscillations are universal and 
reflect the Ramsauer-Townsend effect. 
Figures~\ref{Fig2}~(b) and (e) also show another dramatic effect due to the Q2D 
confinement. The inelastic rates for {\em both} bosonic and fermionic dipoles are 
suppressed as $\exp[-2(d_{\ell}/a_{ho})^{2/5}]$ for values of
$d_{\ell}$ beyond $a_{ho}$, in agreement with Ref.~\cite{Ticknor}.
This result emphasizes the importance of the confinement in allowing for the realization of stable ultracold 
dipolar gases, irrespective of their quantum statistics. 

Figures~\ref{Fig2}~(c) and (f) shows other properties of dipolar scattering through plots of the scattering amplitude versus the 
azimuthal angle, $\phi$, and $d_{\ell}$. 
As one can see, the dipolar scattering is by itself anisotropic. For both bosonic and fermionic dipoles [Figs.~\ref{Fig2}~(c) and (d)], the 
preferential outgoing scattering flux occurs around $\phi=0$, $\pi$, and $2\pi$.
Therefore, two dipoles tend to scatter in such a way that they can either completely repel or else simply "ignore" 
each other. Nevertheless, our results also show that there exist other specific directions in which dipoles can scatter. 
These are indicated by the less intense ``fringes'' between the maximum values of the 
scattering amplitude. 
From a more general perspective, the properties of the scattering amplitude could prove useful
in many-body theories of dipolar gases \cite{ScatAmp}. 

\begin{figure}[htbp]
\includegraphics[width=3.4in,angle=0,clip=true]{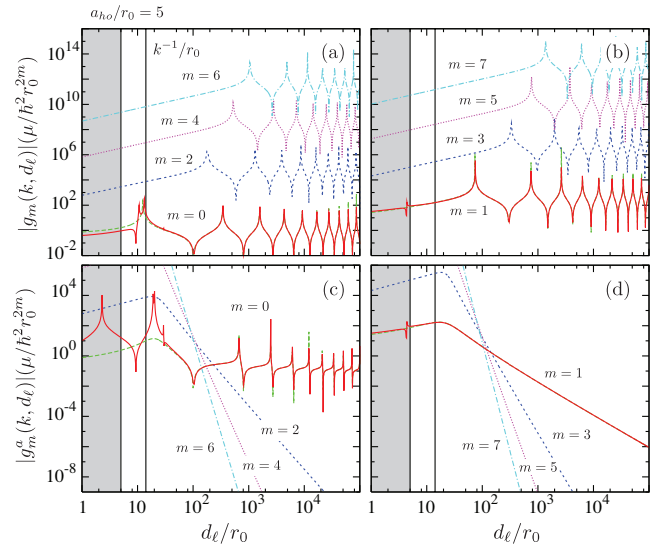}
\caption{(color online). (a) Q2D coupling constant, $g_{m}$, and (b) its equivalent energy-analytic form, $g^{a}_{m}$.
Figure shows interesting ``resonant'' behavior related to the Ramsauer-Townsend (see text).}
\label{Fig3}
\end{figure}

Another way to analyze the many-body behavior is through the pseudopotentials for Q2D dipoles \cite{Q2DPolar1,Q2DPolar2}. 
However, pseudopotentials for dipoles (both in 3D and Q2D) suffer from a number of still-unresolved problems in the limit 
$k\rightarrow0$, and thus their applicability in many-body contexts has remained questionable. Here, we propose
a modified version of Q2D potentials which eliminates the non-analyticity of the pseudopotentials in the limit 
$k\rightarrow0$. Our approach consists in determining the pseudopotentials
with respect to the energy analytic radial 2D solutions in the spirit of generalized quantum defect theory \cite{EnergyAnalytic}.
In the present case this amounts to replacing the usual regular and irregular Bessel solutions by a new pair 
that are both entire analytic functions of energy: $F^{a}_{m}=(k b)^{-m} J_{m}(kr)$ and
$G^{a}_{m}=(k b)^m[Y_{m}(k r)-\frac{2}{\pi}\ln(k b)J_{m}(kr)]$, where $J_{m}$ and $Y_{m}$ are the usual
Bessel functions and $b$ is an appropriate length scale which, based on our numerical calculations, is set to
$b=d_{\ell}$. From this analysis, and following \cite{Q2DPolar1}, we obtain 
the energy-analytic pseudopotentials as,
\begin{align}
&\hat{v}^{a}_{m}(r) \stackrel{r\rightarrow0}{\equiv} g^{a}_{m}(k)\times\delta(r) \hat{R}^{a}_{m}(r),\label{vpp}
\end{align}
where $g^{a}_{m}$ is the energy analytic coupling constant and $\hat{R}^{a}_{m}$ is the regularization operator given, respectively, by
\begin{align}
& g^{a}_{m}=-\frac{\hbar^2}{2\mu}\frac{2}{\pi}\frac{(m!)^2}{(2m)!}(2d_{\ell})^{2m}\tan\delta^{a}_{m}(k),\label{gm}\\
&\hat{R}^{a}_{0}=\ln(\frac{r}{2d_{\ell}\beta_{0}})^2 \frac{d}{dr}\left[\frac{-1}{ \ln(\frac{r}{2d_{\ell}\beta_{0}})}\right],\\
&\hat{R}^{a}_{m}=\frac{1}{r^{m+1}} \frac{d^{2m}}{dr^{2m}}{r^{m}}\left[{1-\frac{2(kr)^{2m}\ln(\frac{r}{2d_{\ell}\beta_{m}})}{2^{2m}m!(m-1)!}}\right]^{-1}\label{op}.
\end{align}
In the equations above, 
$\beta_{m}=e^{-\gamma+\sum_{p=1}^{m}p^{-1}}$,  
where $\gamma\approx0.577$ is the Euler constant 
and $\delta^{a}_{m}$ is the energy 
analytic phase-shift, which relates to the physical phase-shift by 
\begin{eqnarray}
\tan\delta^{a}_{m}(k)=\frac{\tan\delta_{m}(k)/(kd_{\ell})^{2m}}{1-\frac{2}{\pi}(kd_{\ell})^{2m}\ln(kd_{\ell})\tan\delta_{m}(k)}.
\label{tand0}
\end{eqnarray}

Note that our form of the pseudopotential has an explicit dependence on the dipole length $d_{\ell}$ while 
most of the $k$ dependence comes through $\tan\delta_{m}^{a}$. In fact, for $m=0$ our pseudopotential becomes energy
{\em independent} in the $kd_{\ell}\ll 1$ regime.
These properties appear to be absent from previous analyses of Q2D pseudopotentials but they appear 
to be desirable for many-body calculations.
In Figure~\ref{Fig3} we show the contrast between using the usual form of the pseudopotentials and the energy-analytic
form [Eqs.~(\ref{vpp})-(\ref{op})]. Figures~\ref{Fig3}~(a) and (b) shows our results for the physical coupling constant $g_{m}$ 
[obtained by replacing $d_{\ell}$ by $1/k$ and $\delta^{a}_{m}$ by $\delta_{m}$ in Eq.~(\ref{gm})]
for the same set of parameters chosen in Fig.~\ref{Fig2}. As one can see,
$g_{m}$ increases for large $m$ $\tan\delta_{m}/k^{2m}\propto1/k^{2m-1}$ ($m\ne0$),
while $g^a_{m}$ [Figs.~\ref{Fig3}~(c) and (d)] decreases for $kd_{\ell}\gg1$.
This indicates that our form of the pseudopotential might be more suitable in the strongly dipolar regime. 
Apart from the above considerations, we note that for every $m$, $g_{m}$ diverges
at particular values of $d_{\ell}$ \cite{Q2DPolar2}. These divergences are related to the Ramsauer-Townsend
effect and do {\em not} correspond to a zero-energy bound state. If these ``resonances'' persist in 
a many-body context it can indicate that the Q2D dipolar gas could undergo a transition
to a strongly correlated regime where the interactions can be tuned from repulsive ($g_{m}>0$) to 
attractive ($g_{m}<0$) almost at will by varying the external electric field. Moreover, the positions of the 
poles in $g_{m}$, according to our model, should be universal in the regime $k d_{\ell}\gg1$. 
However, since the positions of resonant features depend on $k$ they might be smeared out in a quantum gas. 
But they could be observable in a collision between two clouds at a definite relative momentum \cite{BECCollisions}.
Now, the fact that the poles in $g^{a}_{m}$ are visible only for $m=0$ 
emphasizes the necessity of extending the many-body analysis in the 
energy-analytic framework, a task beyond the scope of our present investigation.

In summary, we have studied scattering properties of bosonic and fermionic dipoles under Q2D confinement.
Both species display universal behavior, which significantly improves the prospects of creating
a stable gas of Q2D dipoles. We have also computed two-body parameters, such as the scattering amplitude and coupling
constants, which can be important for many-body treatments of strongly interacting dipolar gases.

This work was supported by the US-AFOSR-MURI. We thank B. D. Esry, D. Blume, J. L. Bohn, and G. Qu\'em\'ener for
stimulating discussions.

\end{document}